  \documentclass[preprint]{aastex}

\received{2000 February 1}
\accepted{2000 March 1}

\slugcomment{To appear in the {\em Astrophysical Journal Letters.}}

\shorttitle{Accretion Rates onto Black Holes in Quiescent Ellipticals}
\shortauthors{Wrobel \& Herrnstein}
 
\begin{document}
 
\title{Accretion Rates onto Massive Black Holes\\
       in Four Quiescent Elliptical Galaxies}
 
\author{J.~M. Wrobel and J.~R. Herrnstein\altaffilmark{1}}
\affil{National Radio Astronomy Observatory,
       P.O. Box O, Socorro, New Mexico 87801}
\email{jwrobel@nrao.edu, jherrnst@nrao.edu}

\altaffiltext{1}{Current address: Renaissance Technologies 
Corporation, 600 Route 25A, East Setauket, NY 11733-1249}

\begin{abstract}

Four quiescent elliptical galaxies were imaged with the NRAO VLA at
8.5~GHz.  Within the context of canonical advection-dominated
accretion flows (ADAFs), these VLA images plus published black hole
masses constrain the accretion rates to be
$  <1.6\times10^{-4}$,
$  <3.6\times10^{-4}$,
$\le7.8\times10^{-4}$, and
$\le7.4\times10^{-4}$
of the Eddington rates.  These ADAF accretion rates derived at 8.5~GHz
have important implications for the levels of soft and hard X-rays
expected from these quiescent galaxies.

\end{abstract}


\section{MOTIVATION}

Evidence is accumulating that nearby galactic nuclei commonly harbour
massive dark objects \citep{mag98}.  These objects are probably black
holes, because such remnants from the QSO era should be common
\citep{fab95} and because some nuclear star clusters with the
requisite mass and size would be improbably younger that their host
galaxies \citep{mao98}.  Given a black hole mass, $M_{\rm BH}$, a next
important step is to constrain the rate, $\dot{M}$, at which material
is being accreted onto the black hole.  Associated with any black hole
is its Eddington accretion rate, $\dot{M}_{\rm E}$, which is the rate
necessary to sustain the Eddington luminosity.  For a 10\% radiative
efficiency and $M_{\rm BH} = 10^8 m_8~M_{\sun}$, $\dot{M}_{\rm E} =
2.2 m_8~M_{\sun}~{\rm yr}^{-1}$.  For giant elliptical galaxies
hosting black holes, accretion-rate estimates have long been available
for comparison with Eddington rates \citep{fab88,mah97}.  These
estimates are based on \citet{bon52} accretion from an interstellar
medium with temperature $T = 10^7 T_7$~K and pressure $P = 10^6
P_6$~cm$^{-3}$~K, with the accretion being characterized by a Bondi
radius $r_{\rm B} \sim 4.3 m_8 T_7$~pc and a Bondi rate $\dot{M}_{\rm
B} \sim 1.9\times10^{-4} m_8 P_6 T_7 ~M_{\sun}~{\rm yr}^{-1}$.  Then a
black hole with $m_8 = 1$ in a medium with $P_6 =1$ and $T_7 = 1$ will
accrete at a Bondi rate $\dot{M}_{\rm B} \sim
1.9\times10^{-4}~M_{\sun}$~yr$^{-1}$, which is almost four orders of
magnitude less than the associated Eddington rate of $\dot{M}_{\rm E}
= 2.2~M_{\sun}$~yr$^{-1}$.  The Bondi rate estimates are therefore
extremely sub-Eddington.  Still, such low rates could be pervasive
among nearby ellipticals and, moreover, could define a minimum level
of activity for ellipticals hosting massive black holes

This Letter examines the consequences, in the radio and X-ray regimes,
of such low accretion rates in nearby elliptical galaxies.  The
approach is to obtain deep radio continuum images of four ellipticals
studied by \citet{mag98} and then interpret those images within the
context of canonical advection-dominated accretion flows (ADAFs),
reviewed recently by \citet{nar98}.  The radio continuum from an ADAF
is thermal synchrotron emission from a magnetized plasma, and the
conversion between the ADAF radio power, $P_\nu$, and the ADAF
accretion rate, $\dot{M}_{\rm A}$, is relatively straightforward if
the black hole mass is known \citep{mah97,yi98}.  Recent efforts along
these lines have focused on galaxies with significant radio emission
from nonthermal synchrotron jets \citep{dim00}.  In constrast, this
study examines four quiescent elliptical galaxies previously
undetected at radio wavelengths \citep{wro91a,wro91b}, thereby
minimizing one of the largest potential complications in an ADAF
analysis - the role of jet outflows.  Further evidence for the
quiesence of these four galaxies comes from their use as
absorption-line templates in optical spectroscopic studies
\citep{ho97}, plus their weak or undetected X-ray emission
\citep{beu99}.  For these four quiescent ellipticals, the new radio
imaging constrains the ADAF accretion rates to be 
$  <1.6\times10^{-4}$,
$  <3.6\times10^{-4}$,
$\le7.8\times10^{-4}$, and
$\le7.4\times10^{-4}$
of the Eddington rates.  These ADAF accretion rates derived at 8.5~GHz
have important implications for the levels of X-ray emission expected
from these quiescent galaxies.

\section{OBSERVATIONS AND IMAGING}

The Very Large Array (VLA), described by \citet{tho80}, was used 1999
May 6-7 UT in its 1-km configuration to observe NGC\,4291, NGC\,4564,
NGC\,4621, and NGC\,4660.  Data were acquired in dual circular
polarizations and at a center frequency 8.4601~GHz with bandwidth
100~MHz.  Observations were made assuming a coordinate equinox of
2000.  The phase calibrator J1153+8058 was used for NGC\,4291, while
J1239+0730 was used for the other three galaxies.  Scheduled on-source
integration times were 105-120~minutes.  The galaxies were observed at
{\em a priori\/} positions from the Digitized Sky Survey (DSS) web
site.  Improved DSS positions with 1-D error estimates, $\sigma_{\rm
DSS}$, have since become available \citep{cot99} and will be used
below.  $\sigma_{\rm DSS}$ depends in part on the galaxy major axis,
$\theta_{\rm M}$, and values for both terms appear in Table~1.
Observations of 3C\,286 were used to set the amplitude scale to an
accuracy of about 3\%.  The data were calibrated using the 1999
October 15 release of the NRAO AIPS software.  No self-calibrations
were performed.

\placetable{tab1}

The AIPS task IMAGR was used to form and deconvolve images of the
Stokes $IQU\/$ emission from each galaxy.  All images were made with
natural weighting to optimize sensitivity.  The Stokes $I\/$ images
appear in Figure~1, with crosses indicating the the DSS positions and
their errors.  The large ellipses shown in three panels outline the
galaxies' major and minor diameters and elongation position angles
\citep{wro91a}.  The corresponding ellipse for NGC\,4621 falls beyond
the panel boundaries.  For each panel the table gives the image
resolution, $R_{\rm 8.5GHz}$, expressed as the FWHM dimensions and
elongation orientation of the elliptical-Gaussian restoring beam; the
image rms value, $\sigma_{\rm 8.5GHz}$, in units of microjanskys per
beam area ($\mu$Jy~ba$^{-1}$); and the image scale, $s$, based on the
distance, $D$, from Magorrian et al.\ (1998) for
H$_0=80$~km~s$^{-1}$~Mpc$^{-1}$.  Each galaxy image was searched for
an emission peak $S_{\rm 8.5GHz} > 4 \sigma_{\rm 8.5GHz}$ within a
circular region of radius 4 $\sigma_{\rm DSS}$.  The resulting peak
strengths appear in the table.  For the two detected galaxies,
quadratic fits in the image plane yielded the peak flux densities
tabulated, along with their errors that are quadratic sums of a 3\%
scale error and $\sigma_{\rm 8.5GHz}$.  Those fits also yielded the
following radio positions, along with their 1-D errors, $\sigma_{\rm
VLA}$, which are dominated by the peak signal-to-noise ratio
\citep{bal75}: for NGC\,4621,
               $\alpha(J2000) = 12^{h} 42^{m} 02^{s}.50$,
               $\delta(J2000) = 11^{\circ} 38' 48''.6$, and
               $\sigma_{\rm VLA} = 0''.4$;
               for NGC\,4660,
               $\alpha(J2000) = 12^{h} 44^{m} 32^{s}.35$,
               $\delta(J2000) = 11^{\circ} 11' 26''.6$, and
               $\sigma_{\rm VLA} = 0''.7$.
\noindent Each detected galaxy is less than 40\% linearly polarized.

\placefigure{fig1}

\section{IMPLICATIONS}

Radio continuum can be due to thermal synchrotron emission from an
ADAF onto a black hole, nonthermal synchrotron jets emerging from an
ADAF, or nonthermal synchrotron jets emerging from a standard
accretion disk which itself emits no radio photons.  In the radio
regime, this Letter analyzes only thermal synchrotron emission from an
ADAF.  The distances, $D$, from \citet{mag98} are used to convert the
new 8.5-GHz detections or limits to the observed powers, $P_{\rm
8.5GHz}$, given in the table.  These rank among the the most sensitive
radio powers, or power limits, available in the literature for
elliptical galaxies \citep{wro91b}.  Because emission from nonthermal
synchrotron jets could account for some unknown fraction of the radio
detections, the observed peak flux densities will in general impose
upper limits to any ADAF powers, $P_{\rm 8.5GHz}$.

The radio power, $P_\nu$, predicted for an ADAF at a frequency, $\nu$,
depends upon: $\alpha$, the standard Shakura-Sunyaev parametrization
of the kinematic viscosity in the accretion flow; $\beta$, the ratio
of gas pressure to total pressure; $m_8$, the black hole mass in units
of $10^8~M_{\sun}$; $\dot{m}_{\rm A}$, the ADAF accretion rate
normalized to the Eddington rate; and $T_e$, the equilibrium
temperature of the electrons \citep{mah97,yi98}.  Canonical ADAFs
adopt $\alpha=0.3$ and an equipartition $\beta=0.5$.  Moreover, for
most cases of interest $T_e$ is fairly insensitive to the other model
parameters, so an intermediate value of $2.5\times10^9$~K is adopted.
These values for $\alpha$, $\beta$, and $T_e$ then predict that, at
frequency $\nu = 8.5$~GHz,
\begin{equation}
   P_{\rm 8.5GHz} = 6.7\times10^{21} 
                    m_8^{6\over5}
                    \dot{m}_{\rm A}^{6\over5}
                    ~{\rm W~Hz}^{-1} .
\end{equation}
The values for $P_{\rm 8.5GHz}$ from this work and for $m_8$ from
\citet{mag98}, when inserted into Equation~(1), result in the
normalized ADAF accretion rates, $\dot{m}_{\rm A}$, given in the
table.  These normalized ADAF rates are highly sub-Eddington, being
$  <1.6\times10^{-4}$,
$  <3.6\times10^{-4}$,
$\le7.8\times10^{-4}$, and
$\le7.4\times10^{-4}$
of the Eddington rates.  The latter two cases correspond to the
galaxies with radio continuum detections.  Canonical ADAFs exisit only
at normalized accretion rates below a critical rate, $\dot{m}_{\rm
crit} \sim 0.3\alpha^2 = 0.027$ \citep{esi97}.  The normalized rates,
$\dot{m}_{\rm A}$, in the table easily meet this consistency check for
canonical, and other reasonable, values of $\alpha$.  For the four
quiescent ellipticals in this study, the black hole masses from
\citet{mag98} yield the tabulated Eddington accretion rates,
$\dot{M}_{\rm E}$.  Those rates are used to convert the normalized
ADAF rates, $\dot{m}_{\rm A}$, in the table into the absolute ADAF
rates, $\dot{M}_{\rm A}$, also entered in the table.

The ADAF accretion rates derived at 8.5~GHz, in both normalized
($\dot{m}_{\rm A}$) and absolute ($\dot{M}_{\rm A}$) form, have
consequences for the predicted levels of X-ray emission from these
quiescent elliptical galaxies, and these predicted levels must not
violate the observed levels.  Only one galaxy, NGC\,4291, is a weak
X-ray emitter and the other three remain undetected in the ROSAT All
Sky Survey \citep{beu99}.  That survey was conducted with the Position
Sensitive Proportional Counter at soft X-rays (0.5-2 keV) and at an
angular resolution of $\sim 0.5\arcmin$.  The table gives values for,
or upper limits to, the observed soft X-ray luminosities, $L_{\rm
RASS}$, scaled to the \citet{mag98} distances.  No deprojection
analysis has yet been attempted for NGC\,4291, since no ROSAT data
from the High Resolution Imager are available.  Data at hard X-rays
(2-10 keV) are also lacking.

An elliptical galaxy harboring an ADAF will be a source of X-rays from
the ADAF itself and from the galaxy's interstellar medium.  Each
photon source will be discussed in turn.  The normalized ADAF
accretion rates derived at 8.5~GHz are so low that bremsstrahlung
emission, not inverse Compton scattering, will dominate the ADAF's
X-rays \citep{mah97,yi98}.  The latter authors provide an expression
for the bremsstrahlung luminosity of a canonical ADAF that, at the
$T_e$ adopted for Equation~(1), reduces to
\begin{equation}
   L_{\rm A,1keV} = 1.2\times10^{45} m_8 \dot{m}_{\rm A}^2
                    ~{\rm erg~s}^{-1}
\end{equation}
at a fiducial soft X-ray frequency of $\nu = 2.42\times10^{17}$~Hz;
and reduces to
\begin{equation}
   L_{\rm A,6keV} = 7.2\times10^{45} m_8 \dot{m}_{\rm A}^2
                    ~{\rm erg~s}^{-1}
\end{equation}
at a fiducial hard X-ray frequency of $\nu = 1.45\times10^{18}$~Hz.
Applying the tabulated values for $m_8$ and $\dot{m}_{\rm A}$ to
Equations~(2) and (3) yields the tabulated predictions for the soft
and hard X-ray luminosities of the ADAFs.  The predicted luminosities
at 1~keV are consistent with the published ROSAT limits for NGC\,4564,
NGC\,4621, and NGC\,4660, as well as for the ROSAT detection of
NGC\,4291 but only if that detection is dominated by the galaxy's
interstellar medium rather than by its ADAF \citep{beu99}.  Note
further that the ADAFs are expected to be six times more luminous at
hard than at soft X-rays, so these galaxies must must be considered
prime targets for observations in the 2-10~keV region with the
Advanced Satellite for Cosmology and Astrophysics \citep{tan94}.

Soft X-rays can also arise from an elliptical's general interstellar
medium \citep{fab88,mah97,dim00}.  For the galaxies in this study, a
Bondi analysis of that medium can produce an estimate for the Bondi
accretion rate, $\dot{M}_{\rm B}$, which can then be compared with the
absolute ADAF accretion rate, $\dot{M}_{\rm A}$, derived at 8.5~GHz.
The black hole masses from \citet{mag98}, in combination with $T_7 =
1$ being typical for elliptical galaxies \citep{dim00}, results in the
Bondi radii listed in the table.  Assuming further that the pressure,
$P = 10^6 P_6$~cm$^{-3}$~K, at the Bondi radius satisfies $P_6 = 1-10$
\citep{dim00}, then the table gives the corresponding range in Bondi
accretion rates, $\dot{M}_{\rm B}$.  The Bondi rate estimates for $P_6
= 1$ are generally consistent with the limits on the absolute ADAF
rates, $\dot{M}_{\rm A}$, imposed at 8.5~GHz, although there is an
order-of-magnitude discrepancy in the case of NGC\,4291.  In
constrast, for $P_6 = 10$ the Bondi rates exceed the ADAF rates for
all four galaxies.  Meaningful comparions between $\dot{M}_{\rm B}$
and $\dot{M}_{\rm A}$ must clearly await a reliable pressure profile
for ellipticals in general or, better still, for the galaxies in this
study.  Futhermore, integration over any adopted pressure profile
should not violate $L_{\rm RASS}$ listed in the table.

Further radio observations to quantify any jet contamination in
NGC\,4621 and NGC\,4660 are also needed, for two reasons.  First,
removal of jet contamination leads to reduced ADAF emission, which in
turn implies even lower ADAF accretion rates.  Second, the presence of
jet emission could signify the importance of physical processes
ignored in canonical ADAF models and explored by \citet{dim00}.
Follow-up VLA imaging at other frequencies, $\nu$, but at matched
angular resolutions would help assess jet contamination: the flux
density from an ADAF-dominated source is expected to rise as
$\nu^{{1\over3} - {2\over5}}$ \citep{mah97,yi98}, whereas the flux
density from a jet-dominated source is expected to exhibit a flatter
spectral slope.  Although the 500-$\mu$Jy~ba$^{-1}$ upper limits at
5~GHz from \citet{wro91b} do have similar resolutions, they are too
insensitive to impose useful constraints on spectral slopes between 5
and 8.5~GHz.  Moreover, higher-resolution imaging with the NRAO VLBA
could provide morphological evidence for nonthermal synchrotron jets
and spatially separate jet from ADAF emission, as was successfully
done for the spiral galaxy NGC\,4258 \citep{her98}.  Finally, the
image rms values, $\sigma_{\rm 8.5GHz}$, achieved in this study
required 2 hours of integration time.  Similar values for $\sigma_{\rm
8.5GHz}$ will be achieved after a 2-minute integration with the
expanded VLA \citep{nra00}.  This will make it feasible, for the first
time, to target all radio-quiescent ellipticals studied by
\citet{mag98}, which should significantly advance our understanding of
advection-dominated accretion flows as a new class of extragalactic
radio emitters.

\acknowledgments The authors thank Dr.\ R.\ Mahadevan for discussions.
This research has made use of the NASA/IPAC Extragalactic Database
(NED) which is operated by the Jet Propulsion Laboratory, Caltech,
under contract with the National Aeronautics and Space Administration.
The Digitized Sky Surveys were produced at the Space Telescope Science
Institute under U.S. Government grant NAG W-2166. The National
Geographic Society - Palomar Observatory Sky Atlas (POSS-I) was made
by the California Institute of Technology with grants from the
National Geographic Society.  The Second Palomar Observatory Sky
Survey (POSS-II) was made by the California Institute of Technology
with funds from the National Science Foundation, the National
Geographic Society, the Sloan Foundation, the Samuel Oschin
Foundation, and the Eastman Kodak Corporation.  NRAO is a facility of
the National Science Foundation operated under cooperative agreement
by Associated Universities, Inc.

\clearpage

\begin{deluxetable}{llrrrrl}
\footnotesize 
\tablecaption{Quiescent Elliptical Galaxies at 8.5~GHz\label{tab1}} 
\tablewidth{0pc} 
\tablehead{ \colhead{Symbol}   &
            \colhead{Units}    &
            \colhead{NGC\,4291}&
            \colhead{NGC\,4564}&
            \colhead{NGC\,4621}&
            \colhead{NGC\,4660}&
            \colhead{Reference}}
\startdata 
$\theta_{\rm M}$ & $\arcmin$ & 
            2.0&          2.6&          4.5&          2.4& 1\\
$\sigma_{\rm DSS}$ & $\arcsec$ & 
            1.7&          2.0&          3.0&          1.8& 2\\
$R_{\rm 8.5GHz}$ & $\arcsec,\arcsec,\arcdeg$ & 
   15.4,8.4,+56& 10.5,9.0,-22& 10.2,9.0,+19& 14.7,9.1,+54& 3\\
$\sigma_{\rm 8.5GHz}$ & $\mu$Jy~ba$^{-1}$ & 
             16&           13&           13&           14& 3\\
$S_{\rm 8.5GHz}$ & $\mu$Jy~ba$^{-1}$ & 
         $<$ 64&       $<$ 52&   153$\pm$14&   143$\pm$15& 3\\
$D$ & Mpc& 
           28.6&         15.3&         15.3&         15.3& 4\\
$s$ & pc/$\arcsec$& 
            139&           74&           74&           74& 4\\

$P_{\rm 8.5GHz}$ & W~Hz$^{-1}$ & 
   $  <6.3\times10^{18}$& $  <1.5\times10^{18}$& 
   $\le4.3\times10^{18}$& $\le4.0\times10^{18}$& 3\\

$m_8$ & $10^8M_{\sun}$& 
             19&          2.5&          2.8&          2.8& 4\\

$\dot{m}_{\rm A}$ & & 
   $  <1.6\times10^{-4}$&   $<3.6\times10^{-4}$& 
   $\le7.8\times10^{-4}$& $\le7.4\times10^{-4}$& 3\\

$\dot{M}_{\rm E}$ & $M_{\sun}$~yr$^{-1}$ & 
            42.&          5.5&          6.2&          6.2& 4\\
$\dot{M}_{\rm A}$ & $M_{\sun}$~yr$^{-1}$ & 
       $<0.007$&     $<0.002$&   $\le0.005$&   $\le0.005$& 3\\

$L_{\rm RASS}$ & erg~s$^{-1}$ & 
   $ 8.0\times10^{40}$& $<4.2\times10^{39}$& 
   $<4.6\times10^{39}$& $<3.9\times10^{39}$& 5\\

$L_{\rm A,1keV}$ & erg~s$^{-1}$ & 
   $  <5.9\times10^{38}$& $  <4.0\times10^{38}$& 
   $\le2.1\times10^{39}$& $\le1.9\times10^{39}$& 3\\

$L_{\rm A,6keV}$ & erg~s$^{-1}$ & 
   $  <3.5\times10^{39}$& $  <2.3\times10^{39}$& 
   $\le1.2\times10^{40}$& $\le1.1\times10^{40}$& 3\\

$r_{\rm B}$ & pc & 
             82&           11&           12&           12& 4,6\\
$\dot{M}_{\rm B}$ & $M_{\sun}$~yr$^{-1}$ & 
     $0.07-0.7$& $0.001-0.01$& $0.001-0.01$& $0.001-0.01$& 4,6\\
\enddata
\tablerefs{(1) Wrobel 1991; (2) Cotton et al.\ 1999; (3) This work;
(4) Magorrian et al.\ 1998; (5) Beuing et al.\ 1999; (6) Di Matteo et
al.\ 2000.}
\end{deluxetable}

\clearpage

\figcaption[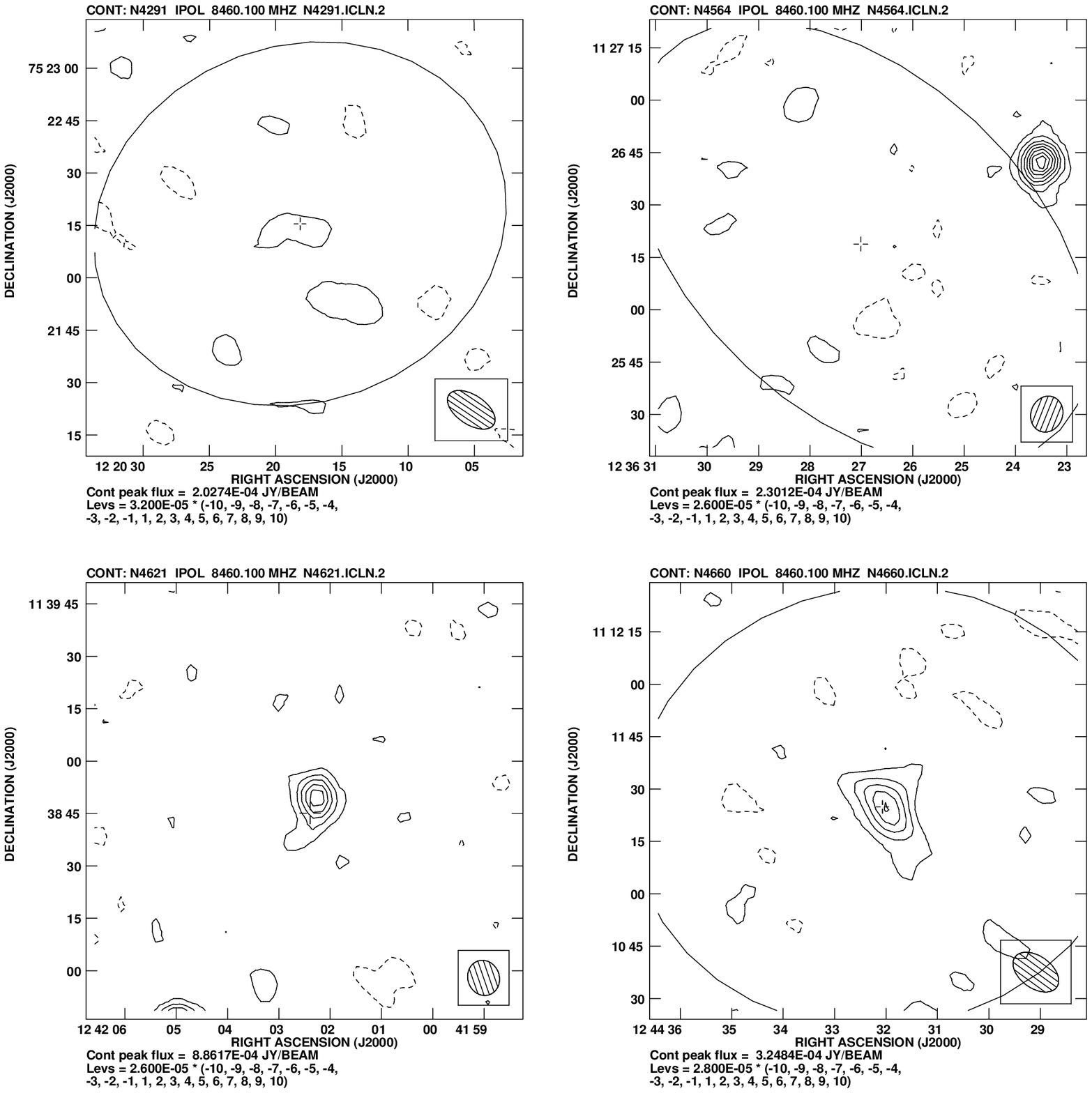]{VLA images of Stokes $I\/$ emission from the inner
$\pm1\arcmin$ of four elliptical galaxies at a frequency of 8.5~GHz.
Contour levels are at multiples of the bottom contour level, which is
twice the image rms value, $\sigma_{\rm 8.5GHz}$.  Hatched ellipses
show the restoring beam areas at FWHM.  Negative contours are dashed
and positive ones are solid.
{\em Upper left:}  NGC\,4291.
{\em Upper right:} NGC\,4564.  
{\em Lower left:}  NGC\,4621.  
{\em Lower right:} NGC\,4660.  
\label{fig1}}

%  Note1 - pgplot files must be edited to move the bounding box to
%  the front from the end.  Otherwise there are lots of problems
%  in plotone.  Same for kntr with grey scale, it seems.
%  Note2 - Montages must be built with divps -E and then bounding
%  boxes aadjusted by hand based on gv to avoid 72 72 72 72.
%
\newpage 
   \epsscale{1.0}
   \plotone{f1.eps}

   \centerline{Figure~1}

\end{document}